\documentclass[entropy,article,accept,moreauthors,12pt,a4paper]{mdpi-arxiv}

\usepackage{graphicx}

\lastpage{x}
\doinum{10.3390/------}
\pubvolume{xx}
\pubyear{2013}
\history{}

\Title{Structural Patterns in Complex Systems Using Multidendrograms}

\Author{Sergio G{\'o}mez $^{1,}$*, Alberto Fern{\'a}ndez $^{2}$, Clara Granell $^{1}$ and Alex Arenas $^{1}$}

\address{%
$^{1}$ Departament d'Enginyeria Inform{\`a}tica i Matem{\`a}tiques, Universitat Rovira i Virgili, Av.\ Pa{\"{\i}}sos Catalans 26, 43007 Tarragona, Spain\\
$^{2}$ Departament d'Enginyeria Qu{\'{\i}}mica, Universitat Rovira i Virgili, Av.\ Pa{\"{\i}}sos Catalans 26, 43007 Tarragona, Spain}

\corres{E-Mail: sergio.gomez@urv.cat.}

\abstract{Complex systems are usually represented as an intricate set of relations between their components forming a complex graph or network. The understanding of their functioning and emergent properties are strongly related to their structural properties. The finding of structural patterns is of utmost importance to reduce the problem of understanding the structure-function relationships. Here we propose the analysis of similarity measures between nodes using hierarchical clustering methods. The discrete nature of the networks usually leads to a small set of different similarity values, making standard hierarchical clustering algorithms ambiguous. We propose the use of \textit{multidendrograms}, an algorithm that computes agglomerative hierarchical clusterings implementing a variable-group technique that solves the non-uniqueness problem found in the standard pair-group algorithm. This problem arises when there are more than two clusters separated by the same maximum similarity (or minimum distance) during the agglomerative process. Forcing binary trees in this case means breaking ties in some way, thus giving rise to different output clusterings depending on the criterion used. Multidendrograms solves this problem grouping more than two clusters at the same time when ties occur.}

\keyword{patterns in networks; hierarchical clustering; dendrogram; uniqueness}

\begin{document}


\section{Introduction}
\label{sec:introduction}

The analysis of structural regularities in the connectivity of complex systems has been a major field of research since the beginning of the so-called ``theory of complex networks'' \cite{newman2003structure,boccaletti06}. The most outstanding approach has been known as community detection, that aims to analyze the mesoscale of complex networks \cite{arenas08,Fortunato201075}. The definition of community in networks is pretty qualitative, a community is a set of nodes more connected between them than with the rest of the network nodes. The success of this approach relies on the findings that have been proved to reproduce some known knowledge about explicit communities in networks. However, less works have concentrated on the finding of other kind of regularities in networks. Particularly interesting is the detection of similarities between nodes \cite{Leicht2006} in a network. Similarities can arise from different measures that represent diverse target patterns, for example, we can search for patterns in local connectivity, or patterns in global connectivity, defining a local distance or global information respectively.
Here we concentrate on the hierarchical clustering of nodes in a network,  based on their connectivity (local/global) similarities.
In general, many pairs of nodes will share the same value of similarity, and then the problem of clustering according to a hierarchy becomes non-unique.

Hierarchical clustering methods are widely used to classify data items into a hierarchy of clusters organized in a tree structure called dendrogram. Agglomerative hierarchical clustering \citep{Gordon1999} starts from a distance matrix between items, each one forming a singleton cluster, and gathers clusters into groups of clusters, the process being repeated until a complete hierarchy of partitions into clusters is formed. There are different types of agglomerative methods such as Single Linkage, Complete Linkage, Unweighted Average, Weighted Average, etc., which only differ by the definition of the distance measure between clusters. To name just a few, uses of hierarchical clustering include the classification of organisms from different populations or species \citep{Sneath1973}, the determination of sets of gens with similar profiles of expression \citep{Eisen1998, DHaeseleer2005}, and the classification of proteins according to sequence similarity \citep{Lazareva-Ulitsky2005, Loewenstein2008}.

Except for the Single Linkage case, all the other agglomerative hierarchical clustering techniques suffer from a non-uniqueness problem, sometimes called the \textit{ties in proximity} problem, when the standard pair-group algorithm is used. This problem arises when there are more than two clusters separated by the same minimum distance during the agglomerative process. The standard approach consists in choosing any pair of clusters, breaking the ties between distances, and proceeds in the same way until a final hierarchical classification is obtained. However, different output clusterings are possible depending on the criterion used to break ties, and very frequently the results of a hierarchical cluster analysis depend on the order of the observations in the input data file.

The ties in proximity problem is well-known from several studies in different fields, for example in biology \citep{Hart1983, Arnau2005}, in psychology \citep{VanDerKloot2005}, or in chemistry \citep{MacCuish2001}. Generally speaking, the problem will arise whenever using discrete values to represent similarity between elements and eventually also with continuous valued functions. The existence of possible ties makes the number of binary dendrograms eventually grow exponentially with the number of elements.
This problem is usually ignored by software packages \citep{Morgan1995, Backeljau1996}, while some other packages just warn against the existence of ties in data sets.

Here we make use of \textit{multidendrograms}, a variable-group algorithm \citep{Fernandez2008} that solves the non-uniqueness problem found in the standard pair-group approach. In Section~\ref{sec:methods} we describe the variable-group algorithm, which groups more than two clusters at the same time when ties occur. In Section~\ref{sec:applications} we show several case studies where we use multidendrograms. Finally, in Section~\ref{sec:conclusions}, we give some concluding remarks.


\section{\textit{Multidendrograms} algorithm}
\label{sec:methods}

Agglomerative hierarchical procedures build a hierarchical classification in a bottom-up way, starting from a distances (or weights) matrix between $n$ individuals. The standard pair-group algorithm has the following steps:
\begin{enumerate}
  \item[0)] Initialize $n$ singleton clusters with one individual in each of them. Initialize also the distances between clusters with the values of the distances between individuals.
  \item[1)] Find the minimum distance separating two different clusters.
  \item[2)] Select two clusters separated by such minimum distance and merge them into a new supercluster.
  \item[3)] Compute the distances\footnote{Depending on the criterion used to compute the distances, different agglomerative hierarchical clusterings are obtained: Single Linkage, Complete Linkage, Unweighted Average, Weighted Average, Unweighted Centroid, Weighted Centroid, and Ward's method are the most commonly used.} between the new supercluster and each of the other clusters.
  \item[4)] If all individuals are not in the same cluster yet, then go back to step~1.
\end{enumerate}

The ties in proximity problem arises when there are more than two clusters separated by the same minimum distance in step~2 of the algorithm. To ensure uniqueness in the agglomerative hierarchical clustering, \textit{multidendrograms} implement a variable-group algorithm \citep{Fernandez2008} that groups more than two clusters at the same time when ties occur. Its main properties are:
\begin{itemize}
  \item When there are no ties, multidendrograms give the same result as the pair-group algorithm.
  \item It always gives a uniquely determined solution thanks to the implementation of the variable-group algorithm.
  \item In the multidendrogram representation of the results, the occurrence of ties during the agglomerative process can be explicitly observed, and a subsequent notion of the degree of heterogeneity inside the tied clusters is obtained.
\end{itemize}

The algorithm has been encapsulated in a public application \textit{MultiDendrograms}~\cite{soft_app} that allows the tuning of many graphical representation parameters, and the results can be easily exported to file. A summary of other of its characteristics are: graphical user interface including data selection, hierarchical clustering options, layout parameters, navigation across the dendrogram, etc.; command-line direct calculation without graphical interface; works both with distances and weights matrices; calculation of ultrametric matrix and deviation measures such as cophenetic correlation coefficient, normalized mean squared error, and normalized mean absolute error; save dendrogram details in text and Newick tree format; and export dendrogram image as JPG, PNG and EPS.


\section{Applications}
\label{sec:applications}

We make use of \textit{MultiDendrograms} to group nodes in networks according to different criteria. In particular, we will show the results of the analysis in three different cases: i) when grouping according to structural vertex similarity, that evaluates common neighbors, ii) when grouping nodes according to their common participation in modules or communities, and iii) when grouping nodes corresponding to a distance matrix dataset.

\subsection{Case study: Vertex similarity in networks}

The idea of classifying vertices in a network according to their similarity in connectivity is a very promising way to unravel missing functional relations or simply to infer missing edges in a certain structure \cite{Holme2005Rolesimilarity,Leicht2006,ChenGZG11}. Imagine, for example, a social network where two nodes have exactly the same acquaintances, it is very likely that these two nodes also present similitudes beyond this specific connectivity profile, that can indicate other common interests. When using the connectivity of the network to asses the similarity of nodes, we usually refer to it as structural similarity. Here we make use of two different definitions, Jaccard and Leicht, provided in the literature to evaluate this similarity. Following \cite{Leicht2006} let us denote by $\Gamma_i$ the set of first neighbors nodes of node $i$, and $|\cdot|$ to denote the cardinality of a given set of elements. The Jaccard similarity \cite{jacc} is defined as
\begin{equation}
\sigma_{ij}^{\mbox{\scriptsize Jaccard}} =\frac{|\Gamma_i\cap\Gamma_j|}{|\Gamma_i\cup\Gamma_j|},
\end{equation}
which encompasses the fraction of common neighbors between two elements. The similarity measure proposed by \citet{Leicht2006} is
\begin{equation}
\sigma_{ij}^{\mbox{\scriptsize Leicht}} =\frac{|\Gamma_i\cap\Gamma_j|}{|\Gamma_i||\Gamma_j|}.
\end{equation}
In this definition the normalization is given by the {\em expected} number of neighbors between any two nodes.

\begin{figure}[!t]
  \begin{center}
  \item[]
  \begin{tabular}[t]{cc}
    \multicolumn{1}{l}{(a)}
    &
    \multicolumn{1}{l}{(b)}
    \\
    \begin{tabular}[b]{c}
    \mbox{\includegraphics*[width=.45\textwidth]{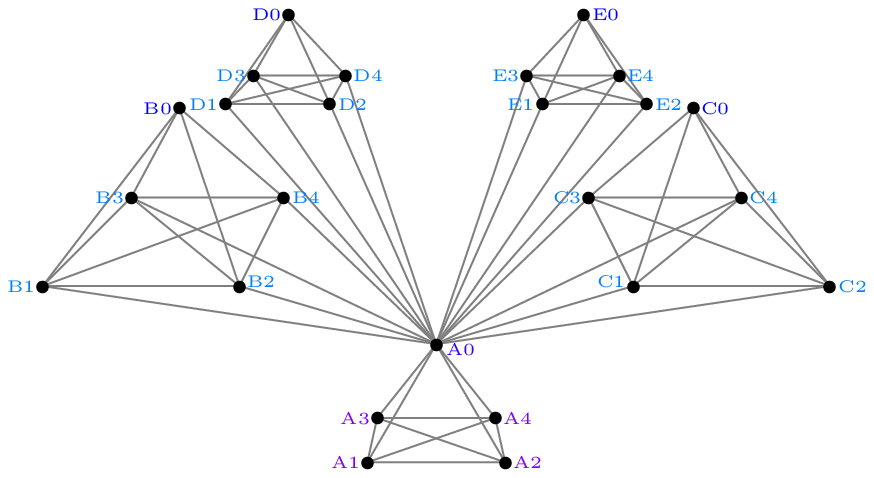}}
    \\ \mbox{\rule{0pt}{30pt}}
    \end{tabular}
    &
    \mbox{\includegraphics*[width=.35\textwidth]{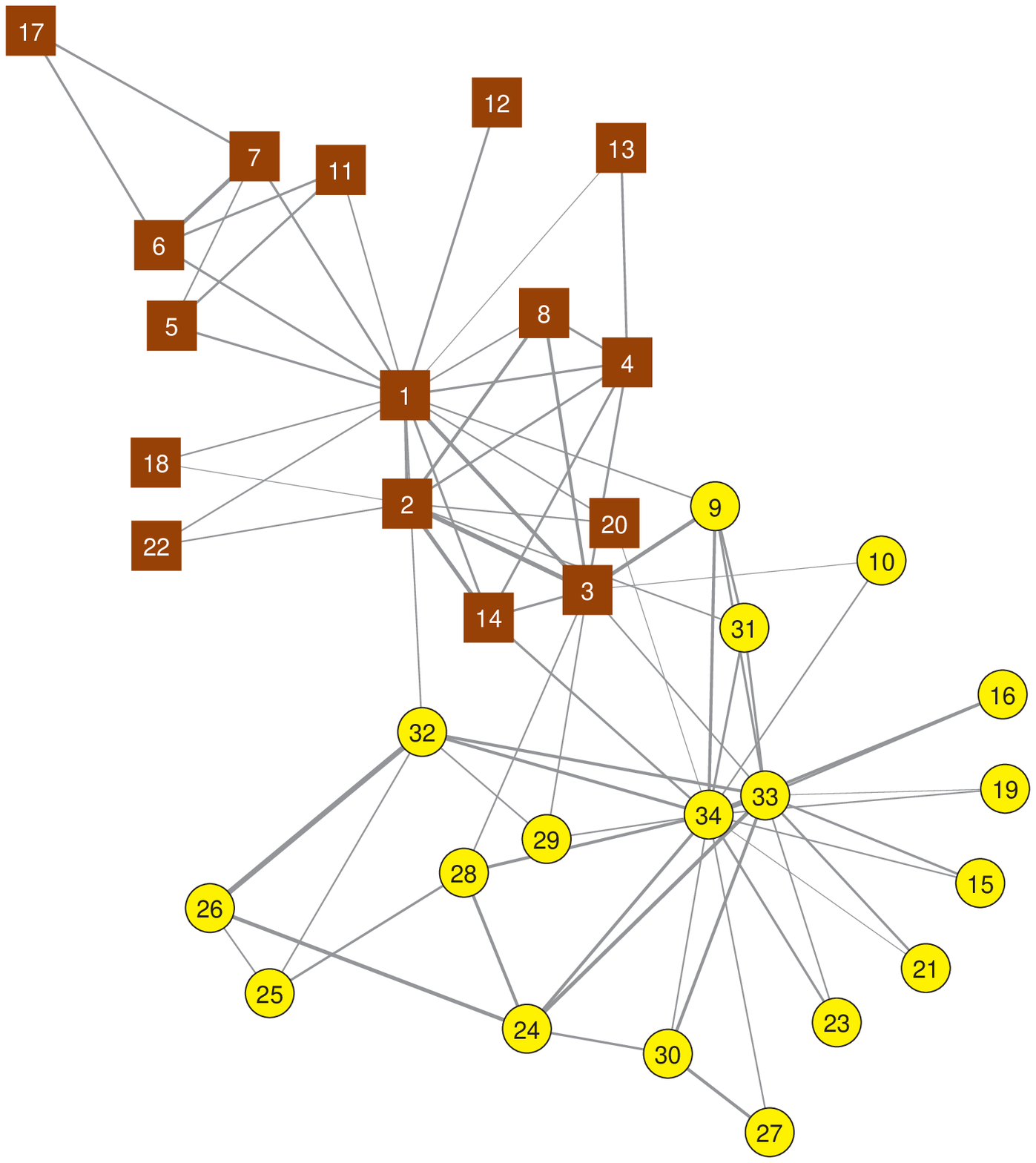}}
    \\
    \begin{tabular}[b]{c}
    \mbox{\includegraphics*[width=.35\textwidth]{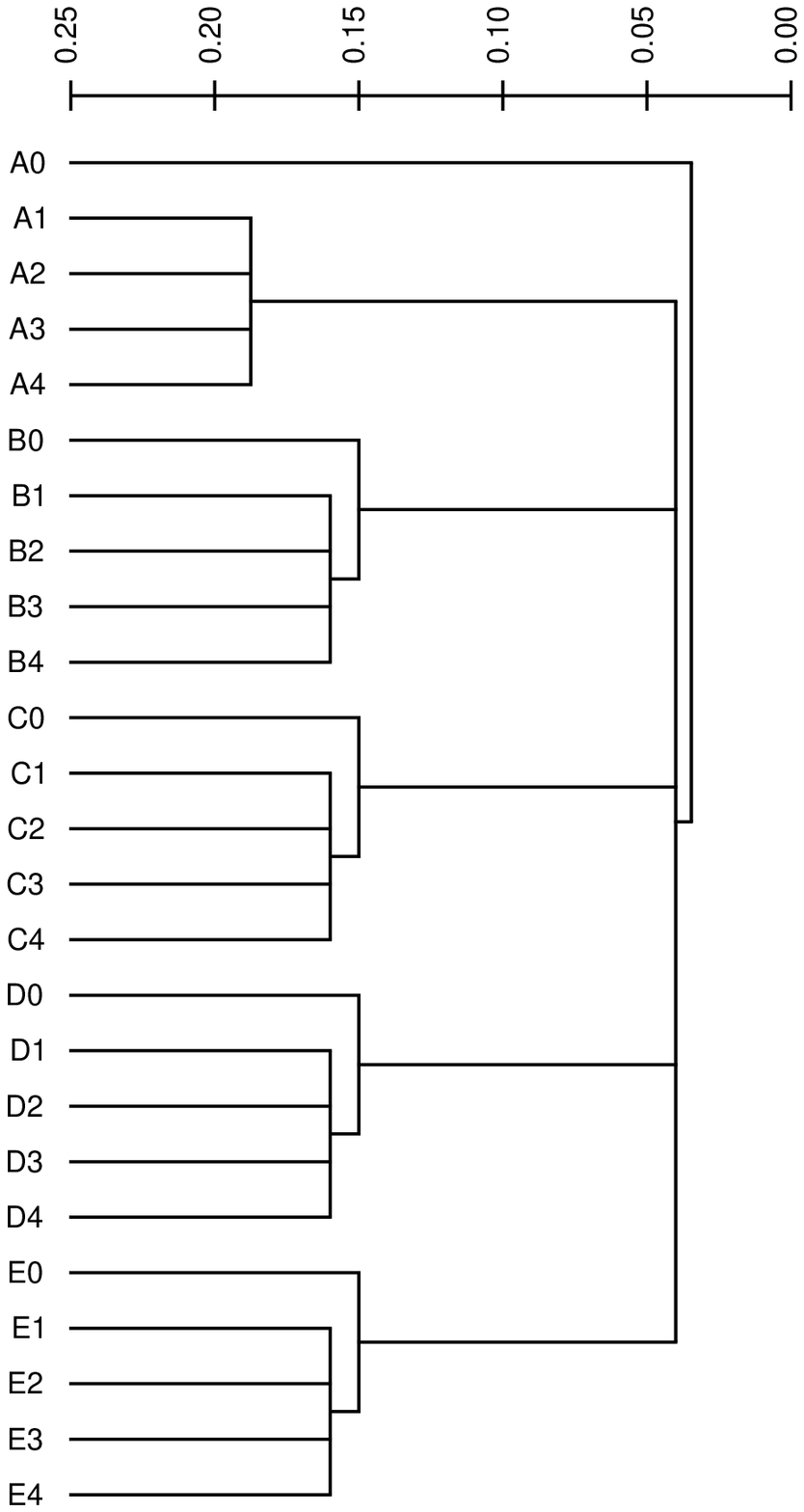}}
    \\ \mbox{\rule{0pt}{18pt}}
    \end{tabular}
    &
    \mbox{\includegraphics*[width=.35\textwidth]{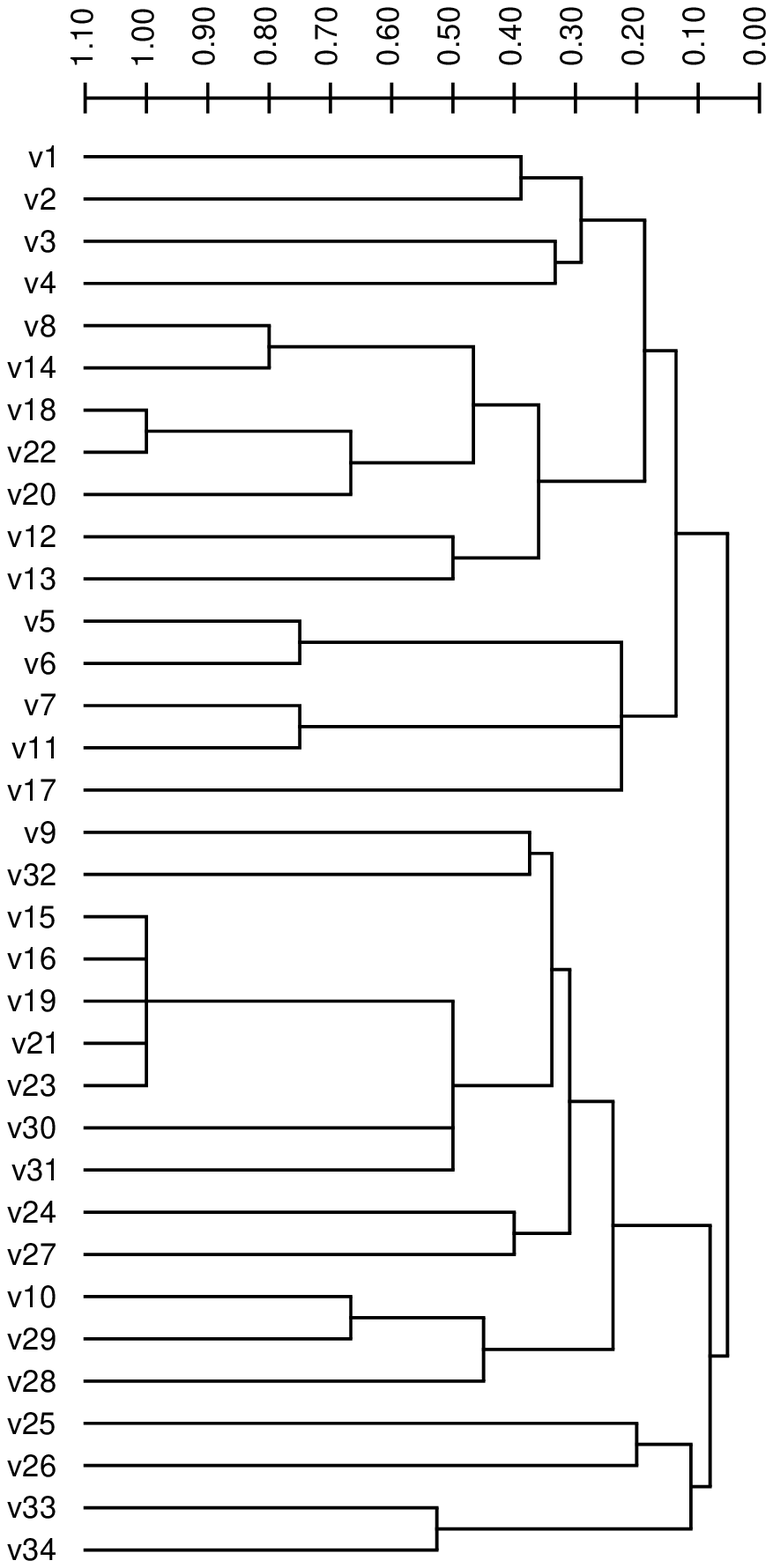}}
  \end{tabular}
  \end{center}
  \caption{(a) Ravasz-Barabasi hierarchical network of 25 nodes. (b) Zachary's karate club network \cite{Zac77}. Multidendrograms obtained using Leicht and Jaccard similarities, respectively. The similarities between clusters are calculated with the standard Unweighted Average method (equivalent to the Unweighted Pair Group Method with Arithmetic Mean, UPGMA, but using variable groups instead of pairs, see \cite{Fernandez2008}).}
  \label{fig1}
\end{figure}

We use a synthetic and a real-world network to apply our algorithm. First, we have investigated the hierarchical scale-free network proposed by Ravasz and Barabasi \cite{rb}. We have implemented a 25 nodes version of the structure whose symmetries are totally pertinent to scrutinize the validity of multidendrograms, Fig.~\ref{fig1}a. Any hierarchical clustering algorithm will differentiate levels in the agglomerative process without having a reason more than pure chance. Multidendrograms, however, correctly identifies the ties and represent them coherently.
Second, we have analyzed the classical social network of the Zachary's karate club \cite{Zac77}, accounting for the study over two years of the friendships between 34 members of a karate club at a US university in 1970. The network in question was divided, at the end of the study period, in two groups after a dispute between the club's administrator and the club's instructor, which ultimately resulted in the instructor leaving and starting a new club, taking about half of the original club's members with him, Fig.~\ref{fig1}b. The analysis of this data has been a paradigmatic benchmark to test the accuracy of community detection algorithms. Note that vertex similarity, which in principle is not a method intended to find communities, finds correctly the real splitting reported in the literature. Moreover, it is easily identifiable the topological symmetries in nodes 15, 16, 19, 21 and 23 that turn out to be a big tie discovered by the multidendrogram.

\subsection{Case study: Modular node similarity in networks}

The fingerprint of modular structure is ubiquitous in real-world complex networks. The standard approach is based on the optimization of a quality function, modularity, which is a relative quality measure for a partition of a network into modules. The analysis of modules at different resolution levels is nowadays a common practice when assessing the significance of the mesoscale of complex networks \cite{arenas08}, and consists in the optimization of the modularity of the graph ${\bf W}_{r}$ for different values of the resolution parameter $r$. Denoting $Q_r$ the modularity of the network at resolution $r$, it reads \cite{arenas08}
\begin{equation}
  Q_r=\sum_{s=1}^{m} \left( \frac{2w_{ss}+n_{s}r}{2w+Nr}-\left(\frac{w_{s}+n_{s}r}{2w+Nr}\right)^{2}\right),
  \label{QWSR}
\end{equation}
where $N$ is the number of nodes, $m$ the number of communities, $w_{ss}$ the internal strength of module $s$, $w_s$ its total strength, $n_s$ its size, and $2w$ the total strength of the network.

The topological scale determined by maximizing $Q$ at which the detection of modular structure has been attacked more frequently corresponds to $r=0$. For positive values of $r$, we have access to the substructures underneath those at $r=0$, and for negative values of $r$ we have access to the superstructures. A detailed analysis can be found in \cite{granell1,granell2}.
The screening of the full mesoscale, i.e.\ the range of values of $r$ for which we detect different modular configurations from individual nodes to the whole network, provides a good representation of the internal structural patterns of the network. However, these patterns vary with the resolution parameter, and it is difficult to assess what is the true structure of the network.
Here, we propose to analyze the mesoscale using a similarity measure for every pair of nodes, which consists in the evaluation of the fraction of the mesoscale that two nodes appear together into the same module. Denoting $V_r$ the set of values of the resolution parameter $r$, and $C_i(r)$ the module at with a certain node $i$ belongs to, we define the modular node similarity as
\begin{equation}
\sigma_{ij}^{\mbox{\scriptsize modular}} =\frac{\displaystyle\left|\sum_r (\delta(C_i(r),C_j(r))\right|}{|V_r|},
\end{equation}
where $\delta$ is the Kronecker function, which takes values 1 when both arguments are equal, and 0 otherwise.

\begin{figure}[!t]
  \begin{center}
  \item[]
  \begin{tabular}[t]{cc}
    \multicolumn{1}{l}{(a)}
    &
    \multicolumn{1}{l}{(b)}
    \\ \\
    \mbox{\includegraphics*[width=.3\textwidth]{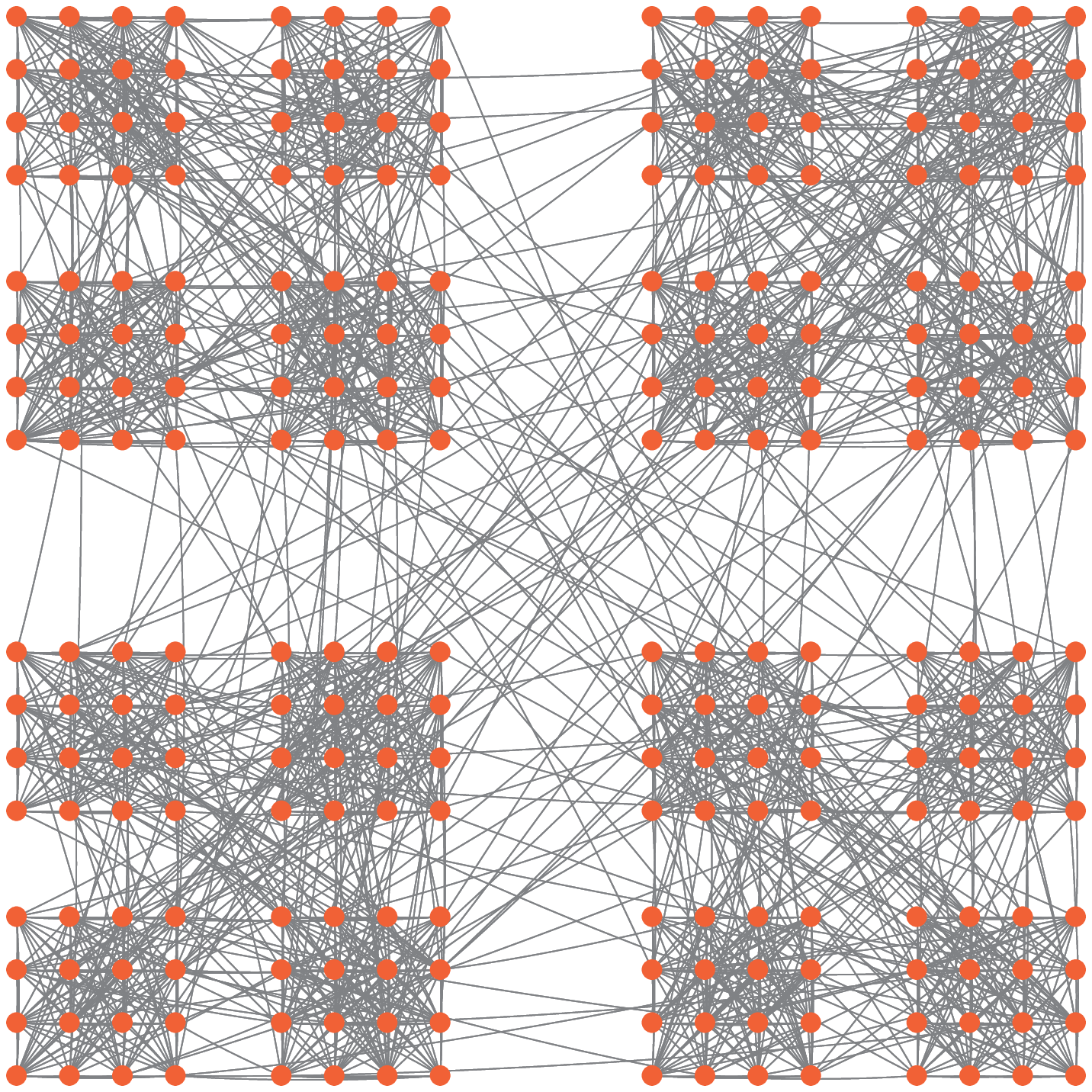}}
    &
    \begin{tabular}[b]{c}
    \mbox{\includegraphics*[width=.49\textwidth]{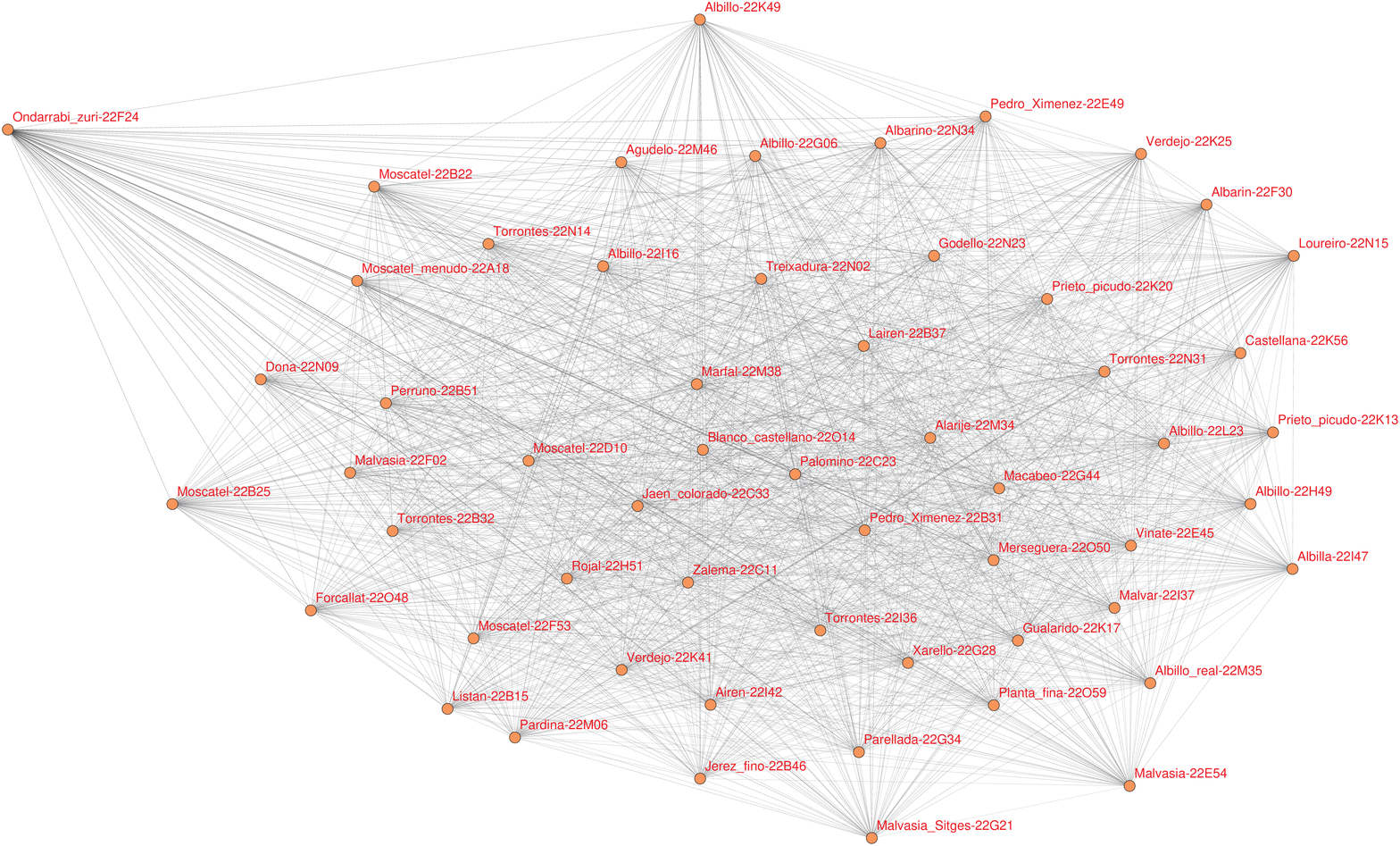}}
    \\ \mbox{\rule{0pt}{10pt}}
    \end{tabular}
    \\ \\
    \mbox{\includegraphics*[width=.35\textwidth]{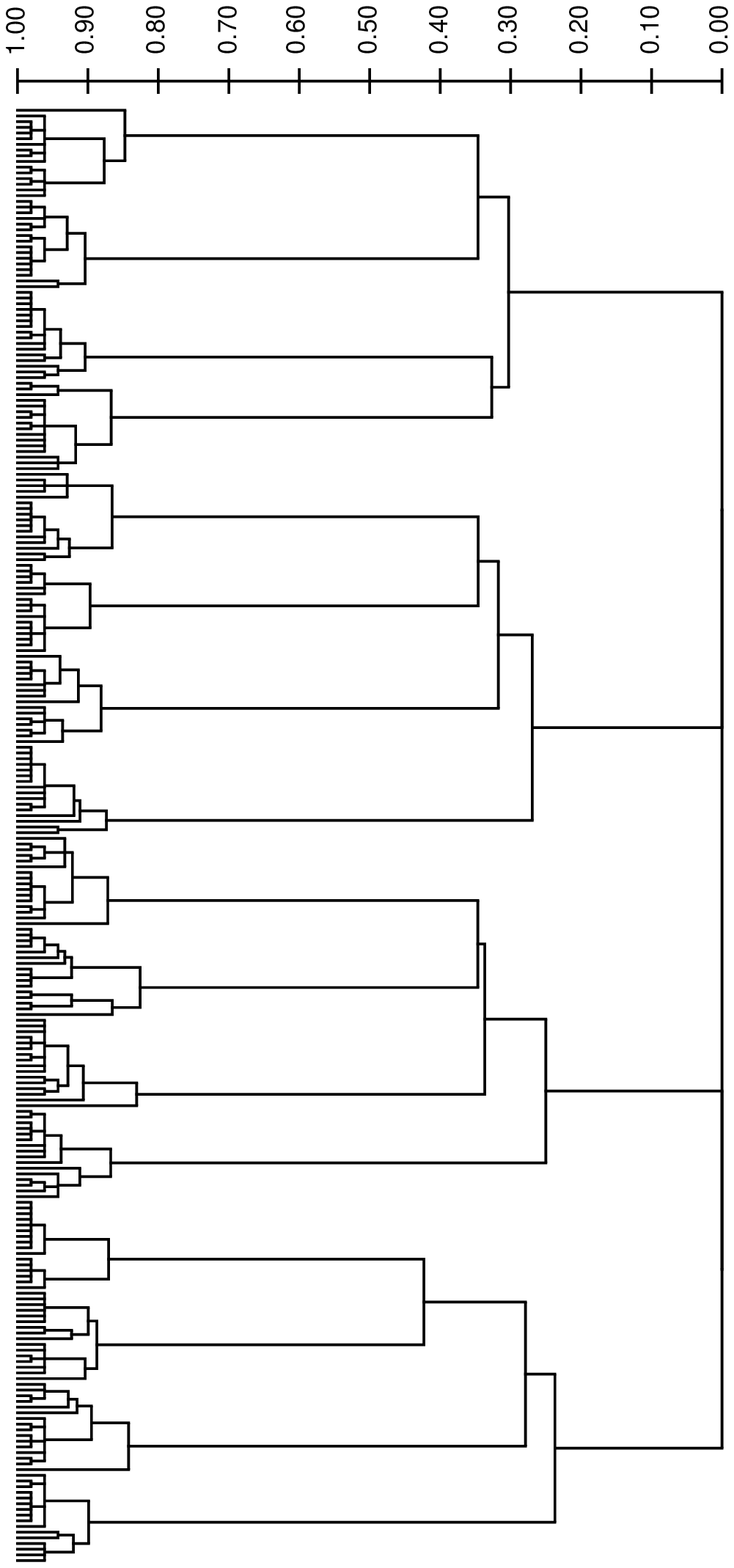}}
    &
    \begin{tabular}[b]{c}
    \mbox{\includegraphics*[width=.43\textwidth]{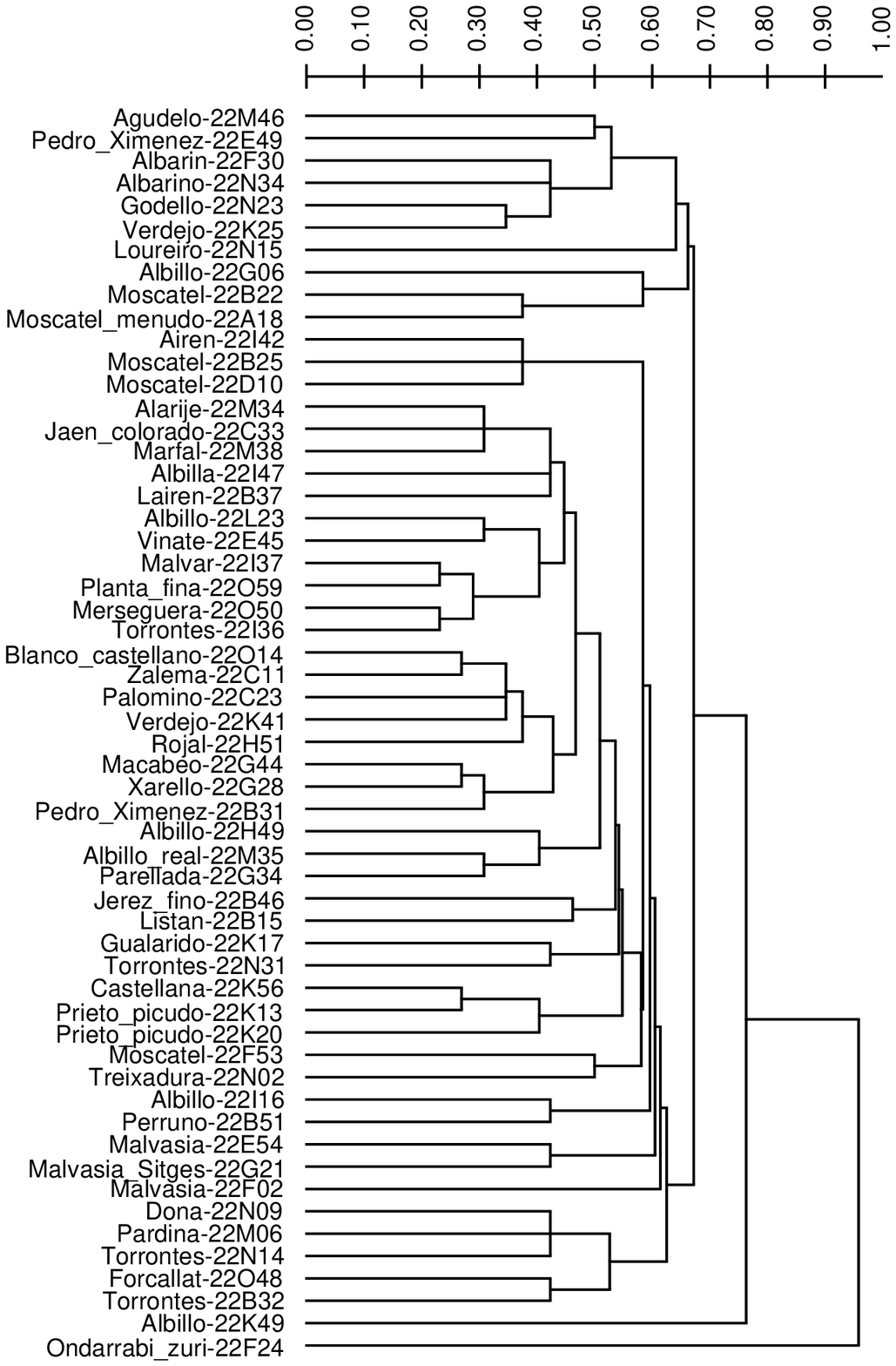}}
    \\ \mbox{\rule{0pt}{23pt}}
    \end{tabular}
  \end{tabular}
  \end{center}
  \caption{(a) H13-4 homogeneous network with modular structure at two different hierarchical levels, and Unweighted Average multidendrogram of its modular similarity. (b) Distances network of white berry varieties in the Spanish grapevine cultivars given in Table~2 of \citet{Ibanez2003}, and its Unweighted Average multidendrogram.}
  \label{fig2}
\end{figure}

We have explored the use of multidendrograms to assess the structure of networks using this similarity. In particular we focus on a synthetic hierarchical network that we call H~13-4 network \cite{Arenas2006}, which corresponds to a homogeneous in degree network with two predefined hierarchical levels, being 256 the number of nodes, 13 the number of links of each node within the most internal community (formed by 16 nodes), 4 the number of links with the most external community (four groups of 64 nodes), and 1 more link with any other node at random in the network.
The results of the multidendrogram reveal the true structure of similarities again in agreement with the underlying hierarchical structured imposed, see Fig.~\ref{fig2}a.
Using the classical binary dendrograms we would obtain several representations none of them capable of join the basic four groups at the same distance.


\subsection{Case study: Distance similarities in complete weighted networks}

Finally we present a more abstract scenario, where the classification is based on a distance matrix between elements. Note that this is equivalent to the analysis of a complete weighted network \cite{jensen}.
Distance ties are very frequent in some types of data such as binary variables, or even integer variables comprising just some few distinct values, because in these cases distances are constrained to a narrow range of values. This is precisely what happens, for instance, in the case of some gene and DNA frequency data \citep{Takezaki1998}, protein-protein interaction data \citep{Arnau2005}, and microsatellite data analysis. In the latter case, biologists usually compute pairwise distances between the multiple locus genotypes of two individuals as a function of the proportion of shared alleles, and the genetic distance matrix obtained is then used to construct a dendrogram, generally with the Unweighted Average clustering method. This is precisely what it is done with the grapevine cultivars data in Table~3 of \citet{Fatahi2003} and the data in Supplementary Table~2 of \citet{Zdunic2013}, where the authors only show one of all possible dendrograms, which amount to $6$ and $36$ respectively. And the problem can grow to unsuspected dimensions such as in the case of the data in Table~2 of \citet{Ibanez2003} and the data in \citet{Almadanim2007}, where the numbers of possible dendrograms are $17\,900$ and $760\,590\,880$ respectively.

Tied distances can also appear using continuous variables, especially if the precision of experimental data is low. Sometimes, on the contrary, the absence of ties might be due to the representation of data with more decimal digits than it should be done. From the family of agglomerative hierarchical methods, Complete Linkage is more susceptible than other methods to encounter ties during the clustering process, since it does not produce new distance values different from the initial ones (Table~\ref{tab:trees}). The non-uniqueness problem also depends on the measure used to obtain the distance values between individuals. Moreover, the larger the data set, the more possibilities for ties to occur.

\begin{table}
  \begin{center}
  \begin{tabular}{lrrr}
    \hline
    Method             & Precision = 3 & Precision = 4 & Precision = 5 \\
    \hline
    Unweighted Average & $17900$ & $2208$ & $2124$ \\
    Weighted Average   & $9859$ & $1709$ & $1762$ \\
    Complete Linkage   & $>10^8$ & $>10^8$ & $>10^8$ \\
    \hline
  \end{tabular}
  \end{center}
  \caption{Number of different binary trees obtained for the grapevine cultivars data in Table~2 of \citet{Ibanez2003}, using distinct hierarchical clustering methods. Although the resolution of the data is equal to 3 significant digits, we show the effect that increasing the precision has on the number of possible binary trees.}
  \label{tab:trees}
\end{table}

In Fig.~\ref{fig2}b we show the distances network corresponding to the subset of 56 white berry varieties in the Spanish grapevine cultivars given in Table~2 of \citet{Ibanez2003}. Genetic distances between genotypes in the nodes were calculated as one minus the proportion of shared alleles at 13 microsatellite loci. In Fig.~\ref{fig2}b we also show the corresponding Unweighted Average multidendrogram for this grapevine network, which has been drawn using a precision equal to the original resolution of the data (i.e., 3~significant digits). In the multidendrogram representation one can clearly observe the occurrence of several ties during the clustering process, which are responsible of up to 509 distinct binary dendrograms. It is also remarkable how the two most dissimilar grapevine varieties, Albillo 22K49 and Ondarrabi zuri 22F24, appear clearly differentiated both in the network representation of the data and in the multidendrogram.



\section{Conclusions}
\label{sec:conclusions}

The search for structural patterns in complex systems is here approached from the hierarchical clustering of its elements according to different similarity (or distance) measures. The correct visualization of the hierarchy is essential to discern these patterns. We have shown the feasibility of the application of multidendrograms to scrutinize the hierarchical structure emerging in different complex networked systems. We have focussed on those representations that because of inherent symmetries will provoke ties when trying to discern which groups to merge. This information can be very useful in several scenarios, e.g.\  discerning groups according to vertex similarity, modular node similarity, and data similarity.

The non-uniqueness problem found in the standard pair-group algorithm for agglomerative hierarchical clustering is usually ignored by the standard algorithms to this end. The software packages ignore or fail to adopt a common standard with respect to ties, many of them simply breaking ties in any arbitrary way. However, different output clusterings are possible depending on the criterion used to break ties, and very frequently the results depend on the input order of the observations. The selection of just one of the possible classifications in such cases can be misleading, and the user is usually unaware of this problem, taking for granted the output given by the software.



\acknowledgements{Acknowledgements}

This work was supported by the European Union (MODERN 309314 to A.F., MULTIPLEX 317532 to S.G. and A.A.); the Spanish Ministry of Science and Innovation (CTM2011-24303 to A.F., FIS2012-38266-C02-01 to S.G. and A.A.); and the Generalitat de Catalunya (2009-SGR-1529 to A.F., 2009-SGR-838 to S.G. and A.A.).
A.A. also acknowledges partial financial support from the ICREA Academia and the James S.\ McDonnell Foundation.


\conflictofinterests{Conflict of Interest}

The authors declare no conflict of interest.


\end{document}